\documentclass[aps,twocolumn,showpacs,superscriptaddress,groupedaddress,nofootinbib]{revtex4} 
\usepackage{graphicx}
\usepackage[sort&compress]{natbib}
\usepackage{subfigure}
\usepackage{amsmath}
\usepackage{amssymb}
\usepackage{amsfonts}
\usepackage{rotating}
\usepackage{cancel}
\usepackage{lipsum}
\usepackage{mathtools}
\usepackage{color}
\usepackage{bbm}
\usepackage{dsfont}
\usepackage{bbold}
\usepackage{multirow}
\usepackage{ulem}

\begin{document}

\title{Sequential single pion production explaining the dibaryon ``$d^*(2380)$''  peak}

\author{R. Molina}
\email{Raquel.Molina@ific.uv.es}
\affiliation{Departamento de F\'{\i}sica Te\'orica and IFIC,
Centro Mixto Universidad de Valencia-CSIC,
Institutos de Investigaci\'on de Paterna, Aptdo. 22085, 46071 Valencia, Spain}   
\author{Natsumi Ikeno}
\email{ikeno@tottori-u.ac.jp}
\affiliation{Departamento de F\'{\i}sica Te\'orica and IFIC,
Centro Mixto Universidad de Valencia-CSIC,
Institutos de Investigaci\'on de Paterna, Aptdo. 22085, 46071 Valencia, Spain}   
\affiliation{Department of Agricultural, Life and Environmental Sciences, Tottori University, Tottori 680-8551, Japan}
\author{Eulogio Oset}
\email{Eulogio.Oset@ific.uv.es}
\affiliation{Departamento de F\'{\i}sica Te\'orica and IFIC,
Centro Mixto Universidad de Valencia-CSIC,
Institutos de Investigaci\'on de Paterna, Aptdo. 22085, 46071 Valencia, Spain}

\begin{abstract}  
We study the two step sequential one pion production mechanism, $np(I=0)\to \pi^-pp$, followed by the fusion reaction $pp\to \pi^+d$, in order to describe the $np\to \pi^+\pi^-d$ reaction with $\pi^+\pi^-$ in $I=0$, where a narrow peak, so far identified with a ``$d(2380)$'' dibaryon, has been observed. We find that the second step $pp\to \pi^+d$ is driven by a triangle singularity that determines the position of the peak of the reaction and the large strength of the cross section. The combined cross section of these two mechanisms produce a narrow peak with the position, width and strength compatible with the experimental observation within the approximations done. This novel interpretation of the peak without invoking a dibaryon explains why the peak is not observed in other reactions where it has been searched for.
\end{abstract}
\maketitle
\section{Introduction}
The $np\to \pi^0\pi^0d$ reaction exhibits a sharp peak around $2370$ MeV with a narrow width of about $70$ MeV, which is also seen in the $pp\to \pi^+\pi^-d$ reaction with approximately double strength \cite{bashkanov,adlarson, adlardos}. In the absence of a conventional reaction mechanism that can explain these peaks, they have been interpreted as a signal of a dibaryon that has been named $d^*(2380)$. On the base of this hypothesis several other features observed in $\pi$ production experiments and $NN$ phase shifts have been interpreted (see \cite{dreview} for a recent review). Actually, the narrow peak in $np\to \pi\pi d$ affects the inelasticity of the $NN$ phase shifts and should have repercussion in $NN$ phase shifts as emphasized in \cite{colin,miguel}. Several mechanisms of two pion production leading to $\pi\pi d$ have been studied in \cite{bashkanov,adlarson,adlardos} based upon the model of \cite{luisal} for $NN\to NN\pi\pi$, which contain double $\Delta$ production, with subsequent $\Delta\to \pi N$ decay or $N^*(1440)$ production with decay of $N^*$ to $N\pi\pi$, or $N^*\to \pi\Delta(N\pi)$. In all these cases the resulting $np$ particles are fused into the deuteron. The results of these calculations give rise to cross sections with small strength compared to the peak of the $np\to \pi^0\pi^0d$ reaction and no peak at the energy of the observed one. Such conclusions were already drawn in an early paper \cite{barnir} and we have explicitly recalculated the cross sections from these mechanisms reconfirming all these earlier findings. Interestingly, in the same work \cite{barnir} a peak with poor statistics, already visible for the $np\to \pi^+\pi^-d$ reaction was explained from a different mechanism, two step sequential $\pi$ production, $np\to pp\pi^-$ followed by $pp\to \pi^+d$. The cross section for $np\to \pi^+\pi^-d$ was evaluated factorizing cross sections for the two latter reactions in an ``on-shell'' approach that called for further checks concerning its accuracy. Such mechanism has no further been invoked concerning the new improved data on the $np\to \pi^+\pi^-d$, $np\to \pi^0\pi^0d$ reactions \cite{bashkanov,adlarson,adlardos}. 
 
\begin{figure}
  \begin{center}
   \includegraphics[scale=0.5]{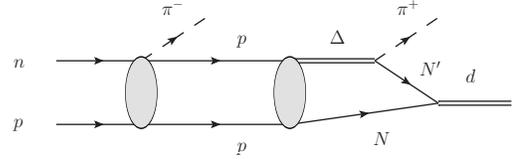}
  \end{center}
\caption{Two step mechanism for $np\to \pi^+\pi^-d$ suggested in \cite{barnir} with explicit $\Delta$ excitation in the $pp\to \pi^+d$ last step as found in \cite{brack, green,weise}. The mechanism with the $nn$ intermediate state is considered in addition.}
\label{fig:dimec}
 \end{figure}

 On the other hand, the time reversal reaction of $pp\to \pi^+d$, $\pi^+$ absorption in the deuteron, $\pi^+d\to pp$, was the subject of study in the past \cite{brack, green,weise} and it was shown to have a neat peak corresponding to the $\Delta$ excitation. Combining the work of \cite{brack, green,weise} with the idea of \cite{barnir} on the $np\to \pi^+\pi^-d$ reaction, the mechanism for $np\to \pi^+\pi^-d$ can be expressed diagrammatically as in Fig. \ref{fig:dimec}.
 
 After many years, more refined data and new theoretical developments make most opportune to revise this issue along the same idea. We can quote:
 
 \begin{itemize}
  \item[1)] The data on $np\to \pi^+\pi^-d$ and $np\to \pi^0\pi^0 d$ have nowadays excellent precision \cite{bashkanov,adlarson,adlardos,dreview}.
  \item[2)] The $np\to \pi^0\pi^0d$ reaction has $\pi^0\pi^0$ in isospin $I=0$, and hence the inital $np$ state must also be in $I=0$.
 
 The work of \cite{adlardos} splits the $np\to \pi^+\pi^-d$ reaction into $I=0$ and $I=1$, and, as expected, the same peak visible in the $np\to \pi^0\pi^0 d$ reaction is seen in the $np\mathrm{(I=0)}\to \pi^+\pi^-d$ reaction with about double strength. This means that in the $np\to \pi^-pp$ reaction, the first step of the sequential single pion production mechanism, the inital $np$ state is also in $I=0$. Only very recently the first step in Fig. \ref{fig:dimec}, $np\to \pi^-pp$ with $np$ in $I=0$ has been singled out with relatively good precission \cite{wasasingle} (see revision about normalization in \cite{wasasingle,tatiana})
 \item[3)] New developments about triangle singularities \cite{landau} allow us to identify the large strength of the $pp\to \pi^+d$ reaction with the presence of a triangle singularity in the triangle diagram shown in the last part of Fig. \ref{fig:dimec}. This corresponds to having simultaneously the $\Delta$ and the two nucleons on shell and collinear. The simplification of the formalism on the triangle singularities done in \cite{bayarguo} allows us to see immediately where the peak of the $pp\to \pi^+d$ cross section should appear, using Eq. (18) of \cite{bayarguo} with the $d$ mass slightly unbound to find a solution of that equation. One predicts that a peak of the cross section should appear around $M_{\mathrm{inv}}(pp)\sim 2179$ MeV, very close to where the peak appears in the experiment \cite{serre}. The Coleman-Norton theorem \cite{coleman} is enlightening in this case to visualize the process. It states that a triangle singularity appears when the process visualized in the triangle diagram can occur at the classical level. In our case this would be: the $pp$ system produces $\Delta N$ back to back in the $pp$ rest frame; the $\Delta$ decays into a $\pi^+$ in the direction of the $\Delta$ and $N'$ in opposite direction, which is the direction of $N$. The $N'$ moves faster than $N$ (encoded in Eq. (18) of \cite{bayarguo}) and catches up with $N$ to fuse into the deuteron. The fusion of the two nucleons into the deuteron comes out naturally when the mechanism discussed has a triangle singularity, giving rise to a neat peak and a cross section rather large compared with typical fusion reactions \cite{dillig}. We have done a recalculation of the $pp\to \pi^+d$ reaction from this new perspective \cite{ikeno}, but the details are unnecessary in the derivation done here for the $np\to\pi^+\pi^-d$ cross section which, as in \cite{barnir}, relies on experimental cross sections, using the new $np \mathrm{(I=0)}\to \pi^-pp$ cross section \cite{wasasingle, tatiana} and the data for $pp\to \pi^+d$ \cite{serre}. We also improve on the on shell approach used in \cite{barnir}. It is also worth mentioning that while the mechanism for $pp\to\pi^+d$ in \cite{brack,weise,green} was not identified as a triangle singularity, it was shown in \cite{green} that the cross section was blowing up when the $\Delta$ width was set to zero, a characteristic of the triangle singularity. In Ref. \cite{ikeno} it is shown that the dominant term in $pp\to \pi^+d$ is the partial wave $^1D_2\,(^{2S+1}L_J)$, in agreement with the experimental observation in \cite{albrow}, and from there one traces back $J^P=1^+,3^+$ for the $d\pi^+\pi^-$ system, with some preference for $3^+$, and $^3D_3$ for the initial $np$ system, the preferred quantum numbers associated to the $d^*(2380)$ peak \cite{dreview}.
 
 It is worth mentioning that the dominance of the $^1D_2$ partial wave leads to a structure suggestive of a resonance in the $pp\to \pi d$ reaction \cite{ueda}, a different dibaryon than the ``$d^*(2380)$''. Theoretical groups also suggest bound states of $N\Delta$, or three body $\pi NN$ \cite{gal} or $N\Delta$ \cite{junhe} to explain the peak of this reaction, although in \cite{huangzhu} they do not find enough binding. Actually, as shown in Ref.~\cite{compass} the Argand plot of a resonance and a triangle singularity are very similar. We stick to the basic rule that if one phenomenon can be explained as conventional, well established facts, this interpretation should be favored against less conventional ones. The works of \cite{brack,green,weise,ikeno} explaining the $pp\to \pi d$ reaction on conventional grounds prove that there is no need of a new dibaryon to explain that reaction.

 \end{itemize}
 \section{Formalism}
 The derivation of the $np\to \pi^+\pi^-d$ cross section that we do follows the steps of the derivation of the optical theorem \cite{itzikson}. We call $t$ the amplitude for isoscalar $np\mathrm{(I=0)}\to \pi^-pp$, $t'$ for $pp\to\pi^+d$ and $t^{\prime\prime}$ for $np\mathrm{(I=0)}\to \pi^+\pi^-d$. The differential cross section for the isoscalar $np\mathrm{(I=0)}\to \pi^-pp$ reaction is given by
 \begin{eqnarray}
  \frac{d\sigma_{np\to \pi^-pp}^I}{dM_\mathrm{inv}(p_1p_1')}=\frac{1}{4ps}(2 M_N)^4\frac{1}{16\pi^3}p_\pi\tilde{p}_1|\bar{t}|^2\frac{1}{2}
  \label{eq:dinp}
 \end{eqnarray}
 where $\sigma^I$ stands for the isoscalar cross section, $\sqrt{s}$ is the center-of-mass (CM) energy of the inital $np$ state, $M_{\mathrm{inv}}(p_1p'_1)$ the invariant mass of the final two protons in this reaction, $p$ the CM momentum of the inital $n$ or $p$ particles, $p_\pi$ the pion momentum in the $np$ rest frame and $\tilde{p}_1$ the momentum of the final protons in the $pp$ rest frame. We use the $(2M_N)^4$ factor of fermion field normalization for the nucleons following the formalism of Mandl and Shaw \cite{mandl}. The magnitude $|\bar{t}|^2$ stands for the angle averaged $|t|^2$ and the factor $\frac{1}{2}$ takes into account the identity of the two final protons. 
 
 Similarly, the cross section for $pp\to \pi^+d$ in the second part of the diagram of Fig. \ref{fig:dimec} is given by 
 \begin{eqnarray}
  \sigma_{pp\to \pi^+d}=\frac{1}{16\pi M_{\mathrm{inv}}^2(p_1p_1')}\frac{p_\pi'}{\tilde{p}_1}|\bar{t}\,'|^2 (2M_N)^2(2M_d)
  \label{eq:dipp}
 \end{eqnarray}
 where $p'_\pi$ is the $\pi^+$ momentum in the $pp$ rest frame and $|\bar{t}'|^2$ stands for the angle averaged $|t'|^2$. We choose to normalize the deuteron field as the nucleons and add the factor $2M_d$ (it disappears from the final formulas).
  On the other hand the amplitude for the $np\to\pi^-\pi^+d$ process in Fig. \ref{fig:dimec} is given by
  \begin{eqnarray}
  -it^{\prime\prime}=&&\frac{1}{2}\int\frac{d^4p_1}{(2\pi)^4}\frac{(2M_N)^2}{2E_N(p_1)2E_N(p'_1)}\frac{i}{p_1^0-E_N(p_1)+i\epsilon}\nonumber\\&&\times\frac{i}{\sqrt{s}-p_1^0-\omega_\pi-E_N(p'_1)+i\epsilon}\,(-i)t\,(-i)t'
  \label{eq:t2}
  \end{eqnarray}
  The factor $\frac{1}{2}$ is to account for the intermediate propagator of two identical particles. In the $d^4p_1$ integrations $t$ and $t'$ would be off shell.  In Ref. \cite{barnir} the pion and the two protons of the intermediate state were taken on shell and $t$ and $t'$ were used with the on shell variables. Theoretical advances done after \cite{barnir} allow us to go beyond this approximation. Indeed, the chiral unitary approach of \cite{npa} for meson-meson interaction, or \cite{angels} for meson-baryon interaction, factorizes the vertices on-shell and performs the loop integral of the two intermediate states. A different justification is given in \cite{ollerulf} writing a dispersion relation for the inverse of the hadron-hadron scattering amplitude, and it also finds a justification in \cite{alberto,kanchan} showing with chiral lagrangians that off shell parts of the amplitudes appearing in the approach get cancelled with counterterms provided with the same theory.  This means that in Eq. (\ref{eq:t2}) we can take $tt'$ outside the $dp_1^0$ integration  with their on-shell values and evaluate the remaining of the integral of Eq. (\ref{eq:t2}). 
  
  Performing the $p_1^0$ integration analytically with Cauchy's residues we get
  \begin{eqnarray}
   t^{\prime\prime}=&&\frac{1}{2}\int\frac{d^3p_1}{(2\pi)^3}\frac{(2M_N)^2}{2E_N(p_1)2E_N(p'_1)}\nonumber\\&&\times\frac{tt'}{\sqrt{s}-E_N(p_1)-E_N(p'_1)-\omega_\pi+i\epsilon}
   \label{eq:t22}
  \end{eqnarray}
where $\vec{p}_1$, $\vec{p}_1\,'$ are the momenta of the intermediate $pp$ particles in Fig. \ref{fig:dimec}, and $\omega_\pi$ the $\pi^-$ energy. The $t,t'$ amplitudes are Lorentz invariant and we choose to evaluate the $\int \frac{d^3p_1}{2E_1(p_1)}$ integral in the $pp$ rest frame, where $|\vec{p}\,'_1|=|\vec{p}_1|$ and $\sqrt{s}-\omega_\pi$ becomes the invariant mass of the two protons. This integral is logarithmically divergent and requires regularization. The result depends smoothly on a cut off $p_{1,\mathrm{max}}$ for $|\vec{p}_1|$ that we use to regularize the $d^3p_1$ integration, and we shall take some values for $p_{1,\mathrm{max}}$ in a reasonable range. Yet, we anticipate that the on shell part given by Eq. (\ref{eq:pro}), below, gives the largest contribution to the $t''$ amplitude. Since $\tilde{p}_1=552$ MeV/c for $M_{inv(p_1p_1')}=2179$ MeV, where the triangle singularity would appear for $t'$ for a $\Delta$ with zero width, or a pronounced peak when the width is considered, values of $p_{1,\mathrm{max}}$ around $700-800$ MeV seem reasonable.

The on-shell approximation used in \cite{barnir} that allows one to write the cross section for $np\to\pi^+\pi^-d$ in terms of the $np\mathrm{(I=0)}\to \pi^-pp$ and $pp\to\pi^+d$ ones is obtained in the present formalism by taking the imaginary part of the two nucleon propagator
\begin{eqnarray}
\hspace{-1cm}&& \frac{1}{M_{\mathrm{inv}}(p_1p_1')-2E_N(p_1)+i\epsilon}\equiv\nonumber\\&& {\cal P}\left[\frac{1}{M_{\mathrm{inv}}(p_1p_1')-2E_N(p_1)}\right]-i\pi\delta(M_{\mathrm{inv}}(p_1p_1')-2E_N(p_1))\nonumber\\
 \label{eq:pro}
\end{eqnarray}
We have then
\begin{eqnarray}
 t^{\prime\prime}_\mathrm{on}=-i\frac{1}{2}\frac{\tilde{p}_1}{8\pi}\frac{(2M_N)^2}{M_\mathrm{inv}(p_1p_1')}\bar{tt'}
 \label{eq:t24}
\end{eqnarray}
where we have factorized the angle averaged value of $tt'$, $\bar{tt'}$.
Using the analogous equation of Eq. (\ref{eq:dinp}) for $d\sigma_{np\to\pi^+\pi^-d}/dM_{\mathrm{inv}}(\pi^+\pi^-)$, the on-shell approximation of Eq. (\ref{eq:t24}) and Eq. (\ref{eq:dinp}) we can already write
\begin{eqnarray}
 \frac{d\sigma_{np\to\pi^+\pi^-d}}{dM_{\mathrm{inv}}(\pi^+\pi^-)}=&&(2M_N)^2(2M_d)p_d\tilde{p}_\pi\frac{1}{4}\frac{\tilde{p}_1^2}{64\pi^2}\nonumber\\&&\times\frac{1}{M^2_\mathrm{inv}(p_1p_1')}\frac{1}{p_\pi\tilde{p}_1}2|\bar{t}'|^2\frac{d\sigma_{np\to\pi^- pp}}{dM_\mathrm{inv}(p_1p_1')}\nonumber\\
 \label{eq:dinpminv}
\end{eqnarray}
where $p_d$ is the deuteron momentum in the original $np$ rest frame, $|\bar{t}'|^2$ the angle averaged $|t'|^2$, and $\tilde{p}_\pi$ the pion momentum in the $\pi^+\pi^-$ rest frame. In Eq. (\ref{eq:dinpminv}) we have assumed that $|\bar{tt'}|^2=|\bar{t}|^2|\bar{t'}^2|$. The amplitudes $t,t'$ in Refs. \cite{wasasingle,serre} have some angular structure, but these are smooth enough to make this assumption a sensible approximation.

Next we use physical arguments to write the $np\to \pi^+\pi^-d$ cross section with an easy compact formula. We note that $\pi^0\pi^0$ or $\pi^+\pi^-$ in $I=0$, as we discussed earlier, require an even value of their relative angular momentum $l$, and when $l=0$ the $\pi^0\pi^0$, or the symmetrized $(\pi^+\pi^-+\pi^-\pi^+)$, behave as identical particles, which reverts into a Bose enhancement when the two pions go together. Certainly if they are exactly together we shall also have the phase space factor $\tilde{p}_\pi$ in $d\sigma/dM_\mathrm{inv}(\pi^+\pi^-)$ of Eq. (\ref{eq:dinpminv}) which makes null this distribution in the two pion threshold, but some enhancement for small invariant masses is expected. Our argumentation is supported by the results of \cite{bashkanov, adlarson} for $\pi^0\pi^0$ (see Fig. 2 of \cite{bashkanov} and Fig. 4 of \cite{adlarson}) and also in \cite{adlardos} for charged pions, although the nature of $I=0$ and $I=1$ in this case distorts a bit the mass distribution compared to the clean $I=0$ $\pi^0\pi^0$ case. 

We could take some $M_\mathrm{inv}(\pi^+\pi^-)$ distribution as input, but to make the results as model independent as possible we take the $\bar{M}_\mathrm{inv}(\pi^+\pi^-)\sim 2m_\pi+60$ MeV, not far from threshold but we change it to see how the results depend on $\bar{M}_\mathrm{inv}$. The stability of the results that we find by changing the value of $\bar{M}_{\mathrm{inv}}(\pi\pi)$ justifies this approximation a posteriori. Then, we can write 
\begin{equation}
 \frac{d\sigma_{np\to\pi^+\pi^-d}}{dM_{\mathrm{inv}(\pi^+\pi^-)}}=\sigma_{np\to\pi^+\pi^-d}\delta(M_{\mathrm{inv}}(\pi^+\pi^-)-\bar{M}_{\pi\pi})\ .
 \label{eq:mpipi}
\end{equation}
The approximation of Eq. (\ref{eq:mpipi}) is sufficiently good and allows us to get a more transparent picture of what is the reason for the appearance of the peak in the $np\to \pi^+\pi^-d$ reaction. 
Note now that the energy of the two pions is obtained as
\begin{equation}
 E_{2\pi}=\frac{s+M^2_\mathrm{inv}(\pi\pi)-M^2_d}{2\sqrt{s}}\ ,
\end{equation}
and since the two pions go relatively together, we take $E_\pi=E_{2\pi}/2$, which allows to relate $M_\mathrm{inv}(p_1p_1')$ with $\sqrt{s}$ via
\begin{eqnarray}
 M^2_\mathrm{inv}(p_1p_1')=(P(np)-p_{\pi^-})^2=s+m^2_\pi-2\sqrt{s}E_\pi
 \label{eq:mppp}
\end{eqnarray}
and formally
\begin{eqnarray}
&& 2M_{\mathrm{inv}}(p_1p_1')dM_\mathrm{inv}(p_1p_1')\nonumber\\&&=-2\sqrt{s}dE_\pi=-M_\mathrm{inv}(\pi\pi)dM_\mathrm{inv}(\pi\pi)\ .
\end{eqnarray}
Using this relationship we can integrate Eq. (\ref{eq:mpipi}) with respect to $M_\mathrm{inv}(\pi\pi)$ and using Eqs. (\ref{eq:dipp}) and (\ref{eq:dinpminv}) we obtain,
\begin{equation}
 \sigma_{np\to\pi^+\pi^-d}=\frac{M_\mathrm{inv}(p_1p_1')}{4\pi}\frac{\sigma_{np\to\pi^-pp}\sigma_{pp\to\pi^+d}}{M_\mathrm{inv}(\pi\pi)}\frac{\tilde{p}_1^2}{p_\pi p'_{\pi}}p_d\tilde{p}_\pi
 \label{eq:dinp2pi}
\end{equation}
One last detail is needed. We have considered the two step $np\mathrm{(I=0)}\to\pi^-pp$ followed by $pp\to\pi^+d$. A properly symmetrized $t^{\prime\prime}$ amplitude requires the addition of $np\mathrm{(I=0)}\to\pi^+nn$ followed by $nn\to \pi^-d$. It is trivial to see considering isospin that the amplitudes $np\mathrm{(I=0)}\to \pi^-pp$ and $np\mathrm{(I=0)}\to \pi^+nn$ are identical up to the phase of $\pi^+$ ($-1$ in our formalism) and the same happens for $pp\to\pi^+d$ and $nn\to \pi^-d$ for the same configuration of the particles. Hence, the product of the amplitudes is the same. In the case that the $\pi^+$ and $\pi^-$ go exactly together, the two amplitudes will be identical and add coherently, but we saw that the phase space factor $\tilde{p}_\pi$ of Eq. (\ref{eq:dinp2pi}) kills this contributions. When considering the integration over the five degrees of freedom of the three body phase space the terms are expected to sum mostly incoherently and we must multiply by $2$ Eq. (\ref{eq:dinp2pi}). Similar arguments can be done with respect to the spin sums and averages. The study of the $pp\to \pi^+d$ reaction in \cite{ikeno} indicates that there is a certain angular dependence on the different spin transitions and we should expect an incoherent sum over spins. Then by including also in $|\bar{t}|^2$ the average over initial spins and sum over final spins we would be considering in our formula the average over spins of the initial $np$ and the sum over spins of the deuteron, plus the intermediate sum over the $pp$ and $nn$ spins.

Eq. (\ref{eq:dinp2pi}) still relies on the on-shell approximation of Eq. (\ref{eq:t24}). To take into account the off shell effects discussed above we realize that factorizing the angular averaged $t,t'$ amplitudes in $t''$ of Eq.~(\ref{eq:t22}), while keeping their energy dependence as a function of $M_\mathrm{inv}(p_1p_1')$, the on shell energy of the intermediate two nucleons, Eq.~(\ref{eq:t22}) has a remaining structure as the $G$ function of two protons,
\begin{eqnarray}
 G=\int\frac{d^3p_1}{(2\pi)^3}\frac{1}{E_N(p_1)E_N(p_1)}\frac{1}{M_\mathrm{inv}(p_1p'_1)-2E_N(p_1)+i\epsilon}\ .\nonumber\\\label{eq:gtwo}
\end{eqnarray}
Then,
\begin{equation}
 \mathrm{Im}G=-\frac{1}{2\pi}\frac{\tilde{p}_1}{M_{\mathrm{inv}}(p_1p_1')}\label{eq:im}\ ,
\end{equation}
where, as mentioned after Eq.~(\ref{eq:dinp}), $\tilde{p}_1$ is the momentum of the two protons in their rest frame.
  The on shell factorization of terms ($t$ and $t'$)  outside the $G$ function was justified in the discussion after Eq.~(\ref{eq:t2}) from different perspectives. We proceed now in a different way to what was done in Eq.~(\ref{eq:t24}) and keep the two terms of Eq.~(\ref{eq:pro}) rather than just the imaginary part. 
In Eq.~(\ref{eq:t24}) we took into account the imaginary part of the integral of Eq.~(\ref{eq:t22}). This is equivalent to taking $\mathrm{Im}\,G$ of Eq.~(\ref{eq:im}) instead of $G$ in the integral of Eq.~(\ref{eq:gtwo}). To revert this approximation and find the effects of the off shell part of the integral approximately we replace in Eq.~(\ref{eq:dinp2pi}),
\begin{eqnarray}
 \left(\frac{\tilde{p}_1}{2\pi M_\mathrm{inv}(p_1p_1')}\right)^2\to |G(M_\mathrm{inv})|^2\ .\label{eq:gminv}
\end{eqnarray}

The last step in the evaluation of $\sigma_{np\to\pi^+\pi^-d}$ requires to use the experimental data for $np(I=0)\to\pi^-pp$ and $pp\to \pi^+d$. We get $\sigma_{pp\to \pi^+d}$ directly from experiment \cite{serre}. The $\sigma_{np\to\pi^-pp}$ in $I=0$ requires some thoughts. In \cite{dakhno,bystricky} the isoscalar $NN\to\pi NN$ amplitude is obtained via isospin symmetry from $\sigma_{np\to pp\pi^-}$ and $\sigma_{pp\to pp\pi^0}$, and relatively precise results are obtained in \cite{wasasingle} from improved measurements of these cross sections. In the erratum of \cite{wasasingle} and in \cite{tatiana} it is clarified that the actual $\sigma_{pn(I=0)\to NN\pi}$ is one half of $\sigma_{NN(I=0)\to NN\pi}$ of \cite{wasasingle}. The cross section that we need is $\sigma_{pn(I=0)\to pp\pi^-}$. It is trivial to see using isospin symmetry that $\sigma_{pn(I=0)\to pp\pi^-}$, $\sigma_{pn(I=0)\to nn\pi^+}$ and $(\sigma_{pn(I=0)\to pn\pi^0}+\sigma_{pn(I=0)\to np\pi^0 })$ are all equal. Then we write the relationship of \cite{dakhno,bystricky,wasasingle} as 
\begin{eqnarray}
\sigma_{np(I=0)\to pp\pi^-}= &&\frac{1}{3}\sigma_{np(I=0)\to NN\pi}\label{eq:isorel}\\&&=
 \frac{1}{6}\sigma_{NN(I=0)\to NN\pi}\nonumber\\&&=\frac{1}{6}3(2\sigma_{np\to pp \pi^-}-\sigma_{pp\to pp\pi^0})\nonumber
\end{eqnarray}
and we take data for $\sigma_{np(I=0)\to NN\pi}$ from Fig. 1 of \cite{tatiana}. Statistical and some systematic errors are considered in \cite{wasasingle,tatiana}. With the only purpose of making a realistic fit to the data we include also systematic errors from the uncertainty in Eq. (\ref{eq:isorel}) when using isospin symmetry. We assume a typical $5\%$ violation of isospin in each of the last two terms of Eq. (\ref{eq:isorel}) and sum the errors in quadrature. The systematic errors obtained are of the order of $0.5$ mb in $\sigma_{np(I=0)\to NN\pi}$, which we also add in quadrature to the former ones of \cite{tatiana}. With those errors we have many good fits with reduced $\chi^2$, ($\chi^2_r$), smaller than $1$. We take two of them, one peaking on the lower side of $\sqrt{s}$ and the other one on the upper side for the $np\mathrm{(I=0)}\to NN\pi$ cross section, parameterized as,
\begin{eqnarray}
 \sigma_i=\left|\frac{\alpha_i}{\sqrt{s}-\tilde{M}_i+i\frac{\tilde{\Gamma}}{2}}\right|^2
 \label{eq:sig}
\end{eqnarray}
and call set I the one with the parameters: $\tilde{M}_1=2326$ MeV, $\tilde{\Gamma}_1=70$ MeV, $\alpha^2_1=2.6\left(\frac{\tilde{\Gamma}_1}{2}\right)^2$ mb MeV$^2$ ($\chi^2_r=0.50$), while set II has $\tilde{M}_2=2335$ MeV, $\tilde{\Gamma}_2=80$ MeV, $\alpha^2_2=2.5\left(\frac{\tilde{\Gamma}_2}{2}\right)^2$ mb MeV$^2$ ($\chi^2_r=0.52$)\footnote{One should not attribute this shape to the Roper excitation as assumed in \cite{wasasingle,tatiana}. We have seen that the Roper excitation grows smoothly monotonically around this energy region (see also Fig. 1 of Ref. \cite{tatiana}) and there are many other mechanisms contributing to the amplitude with cancellations among them.}. The $pp\to\pi^+d$ cross section has accurate data and we parameterize it as 
\begin{eqnarray}
 \sigma_3=\left|\frac{\alpha_3}{M_\mathrm{inv}(p_1p'_1)-\tilde{M}_3+i\frac{\tilde{\Gamma}_3}{2}}\right|^2
 \label{eq:sig1}
\end{eqnarray}
with $\tilde{M}_3=2165$ MeV, $\tilde{\Gamma}_3=123.27$ MeV, $\alpha^2_3=3.186\left(\frac{\tilde{\Gamma}_3}{2}\right)^2$ mb MeV$^2$.

\begin{figure*}
 \begin{center}
 \begin{tabular}{cc}
  \includegraphics[scale=0.4]{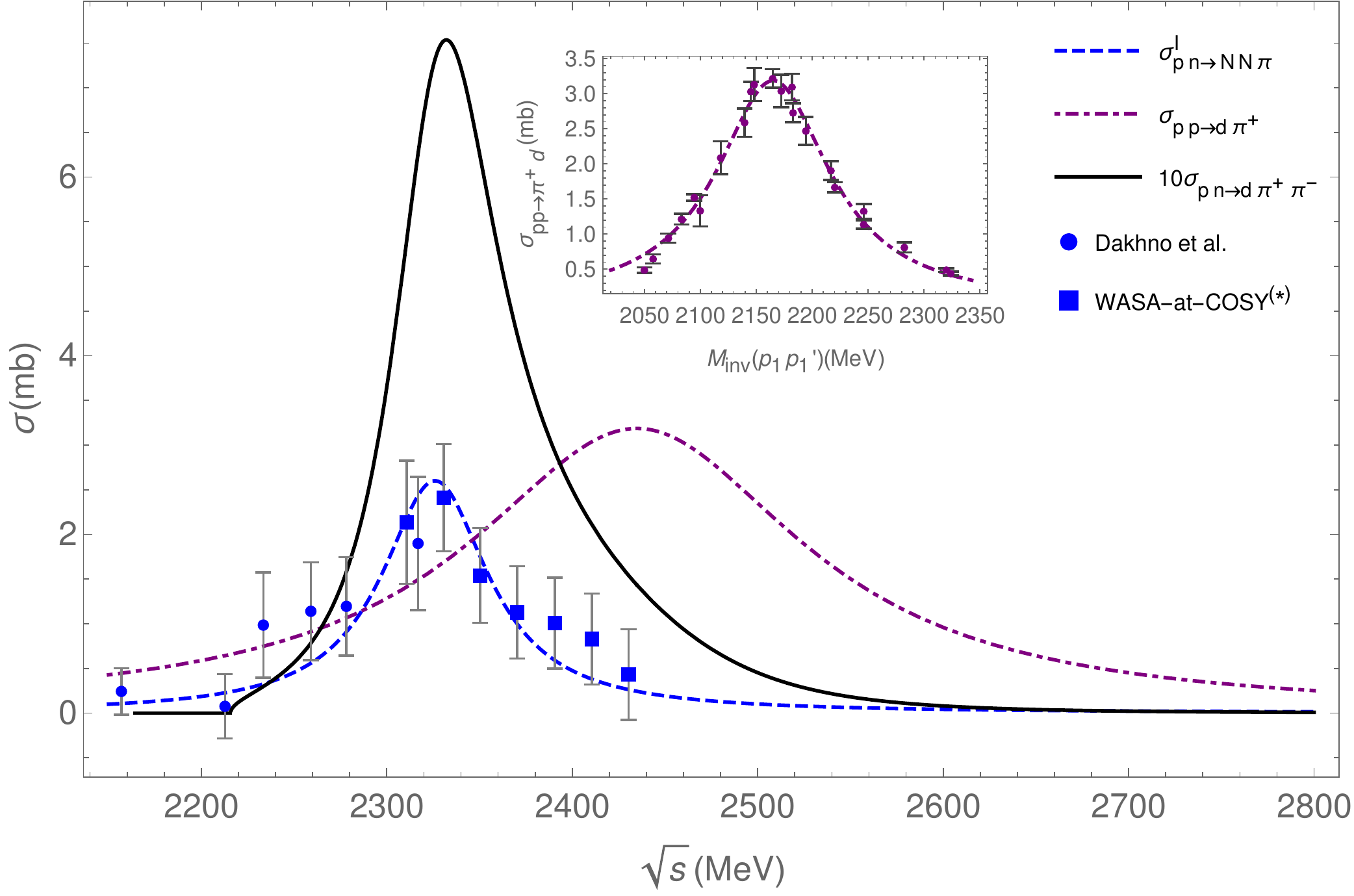}&\includegraphics[scale=0.4]{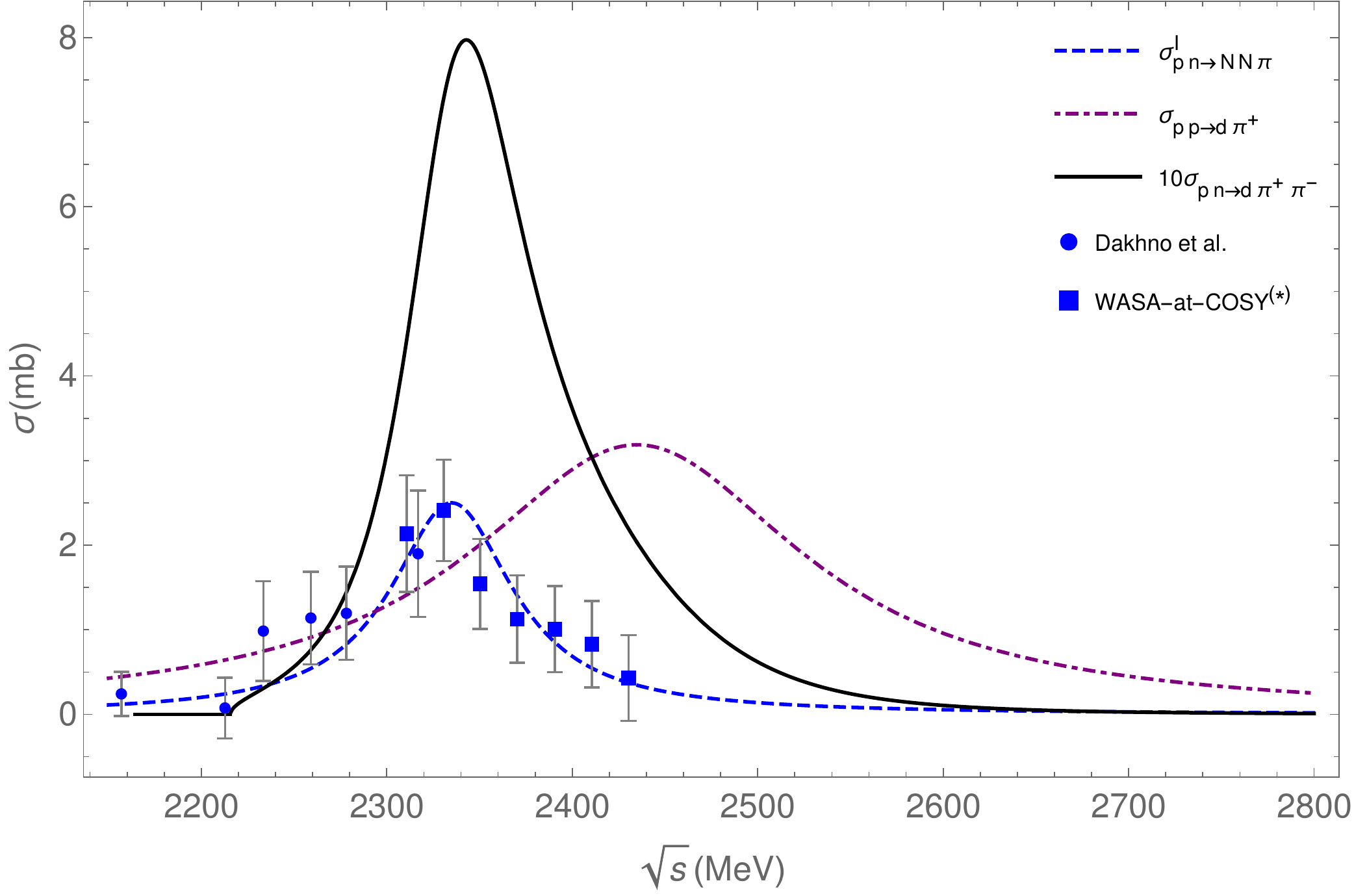}\\
 \end{tabular}
 \end{center}
 \caption{Plots of $\sigma_{np\to\pi^-pp}(I=0)$ and $\sigma_{pp\to\pi^+d}$, as a function of $\sqrt{s}$ and $M_{\mathrm{inv}}(p_1p_1')$, respectively, where $M_{\mathrm{inv}}(p_1p_1')$ is evaluated by means of Eq. (\ref{eq:mppp}). The results with $\sigma_{np\to\pi^+\pi^-d}$ in $I=0$ of Eq. (\ref{eq:sigtot}) are multiplied by $10$ for a better comparison. Left: Results with set I; Right: Results for set II. $\bar{M}_{\pi\pi}=2m_\pi+60$ MeV. Inset: $\sigma_{pp\to\pi^+d}$ as a function of $M_{\mathrm{inv}}(p_1p'_1)$. Data for $pp\to\pi^+d$ from \cite{serre}. Data for $np\mathrm{(I=0)}\to\pi NN$ are taken from Dakhno et al. \cite{dakhno}, and WASA-at-COSY$^{(*)}$ \cite{wasasingle,tatiana}, including systematic errors from isospin violation.}
\label{fig:compa}
\end{figure*}

With the former discussions our final formula on shell is given by
\begin{eqnarray}
 \sigma_{np\to\pi^+\pi^-d}=\frac{M_{\mathrm{inv}}(p_1p'_1)}{6\pi}\frac{\sigma^I_{np\to  NN\pi}\sigma_{pp\to\pi^+d}}{M_{\mathrm{inv}}(\pi\pi)}\frac{\tilde{p}_1^2}{p_\pi p'_\pi}p_d\tilde{p}_\pi\nonumber\\
 \label{eq:sigtot}
\end{eqnarray}
with $\sigma^I_{np\to NN\pi}=\sigma_{np(I=0)\to NN\pi}$ of \cite{wasasingle,tatiana} and we show the results in Fig. \ref{fig:compa}.

We can see that the cross section of $pp\to \pi^+d$ and $np\mathrm{(I=0)}\to NN\pi$ overlap around the middle of their energy distributions such that their product in Eq. (\ref{eq:sigtot}) gives rise to a narrow peak around $\sqrt{s}=2340$ MeV, close to the position of the experimental $np\to\pi^+\pi^-d$ peak around $2365$ MeV.

\begin{table}
\caption{Values of the peak strength (``strength"), peak position (``position"), and width, for intermediate particles on shell (columns with $\delta \bar{M}_{\pi\pi}$, $\bar{M}_{\pi\pi}=2m_\pi+\delta \bar{M}_{\pi\pi}$), and off-shell (``o.s."), where we have taken $p_{1,\mathrm{max}}=700$ and $800$ MeV and $\delta\bar{M}_{\pi\pi}=60$ MeV.} 
 \begin{center}
 \setlength{\tabcolsep}{0.5em}{\renewcommand{\arraystretch}{1.7} 
 \begin{tabular}{l|r|r|r|r|r}
&\multicolumn{3}{c|}{$\delta\bar{M}_{\pi\pi}$ (MeV)}&\multicolumn{2}{c}{$p_{1,\mathrm{max}}^{\mathrm{o.s.}}$ (MeV)}\\
\hline
Set I&$40$ &$60$ &$80$&$700$&$800$\\
strength (mb)&$0.72$&$0.76$&$0.75$&$0.82$&$0.95$\\
position (MeV)&$2332$&$2332$&$2332$&$2332$&$2332$\\
width (MeV)&$76$&$76$&$81$&$75$&$75$\\
Set II&&&&&\\
\hline
strength (mb)&$0.75$&$0.80$&$0.80$&$0.85$&$0.96$\\
position (MeV)&$2342$&$2345$&$2345$&$2343$&$2342$\\
width (MeV)&$86$&$87$&$88$&$87$&$84$\\
\hline
 \end{tabular}}

 \end{center}
\label{tab:para}
\end{table}

In table \ref{tab:para} we show the results  obtained with set I and set II for the strength of $\sigma_{np\to\pi^+\pi^-d}$ at the peak, the peak position and the width of the peak, varying $\bar{M}_{\pi\pi}$, and $p_{1,\mathrm{max}}$ for the off shell calculations. What one sees is a stability of the results upon changes of $\bar{M}_{\pi\pi}$, which justifies the use of Eq. (\ref{eq:mpipi}). We also find that off shell effects using Eq. (\ref{eq:gminv}) are small, justifying the on shell approximation used in \cite{barnir}. The strength at the peak between $0.72-0.96$ mb should be considered quite good compared to the experimental one around $0.5$ mb, given the different approximations done (fits to the $np\mathrm{(I=0)}\to \pi^-pp$ cross section with the systematic errors with $20-30$ \% smaller strength at the peak are still acceptable, hence such uncertainties in the resulting $np\to\pi^+\pi^-d$ cross section are expected). The peak position from $2332-2345$ MeV should also be considered rather good compared to the about $2365$ MeV of the experiment \cite{bashkanov,adlarson,adlardos,wasasingle}. The narrow width observed in the experiment of $70-75$ MeV is also well reproduced by our results in the range of $[75-88]$ MeV. 

The appeareance of the peak about $25$ MeV below the experimental one is not significant with the perspective that, as discussed in \cite{wasasingle}, the authors achieve a resolution in $\sqrt{s}$ of about $20$ MeV and the $pp\to pp\pi^0$ and $pn\to pp\pi^-$ cross sections, from where $\sigma_{np\mathrm{(I=0)}\to pp\pi^-}$ is obtained via Eq. (\ref{eq:isorel}) with large cancellations, are measured using data bins of $50$ MeV in $T_p$.

The derivation done contains the basic dynamical ingredients in a skilled way, making some approximations to rely upon experimental cross sections. We think that it is remarkable that a narrow peak, at about the right position, with strength and width comparable to the experimental peak of $np\to\pi^+\pi^-d$, appears in spite of the approximations done, and the stability of the results allows us to conclude that a peak with the properties of the experimental one associated so far to the ``$d^*(2380)$" dibaryon is unavoidable from the mechanism that we have studied.

From the perspective of the $np\to\pi^+\pi^-d$ reaction being tied to the particular reaction mechanism of Fig. \ref{fig:dimec}, with a two step sequential one pion production, it is easy to understand why the narrow peak of the $np\to \pi^+\pi^-d$ reaction is not seen in $\gamma d\to \pi^0\pi^0d$ in spite of having the same final state \cite{guenther}. The first reaction is a fusion reaction, with the last step tied to a triangle singularity. The $\gamma d\to \pi^0\pi^0d$ reaction is a coherent reaction, the $d$ is already present in the initial state and the reaction mechanisms are drastically different. 

\section{Conclusions}
In summary, we have identified the reaction mechanism that produces a narrow peak in the $np\to \pi^+\pi^-d$ cross sections without having to invoke a ``dibaryon" resonance. From this perspective it is also easy to understand why the peak is not seen in other reactions where it has been searched for, although the peak contributing to the inelastic channels of $pn\to all$ can show traces in $NN$ phase shifts, as anticipated in \cite{colin, miguel} and discussed in \cite{dreview}.

\section{Discussion about other reactions claiming to see the ''$d^*(2380)$`` state}
The $d^*(2380)$ has been claimed to be seen in other reactions (see review in \cite{clementchin}). At the same time the results of the former section, posted in arXiv (arXiv: 2102.05575) have been scrutinized in Ref.~\cite{comment} where apparent contradictions with experiment have been claimed. In what follows we address the different points raised in Ref.~\cite{comment} to show that there are no contradictions of our approach with experiment and that the ``proofs'' presented in favor of the dibaryon hypothesis are unfounded. We follow the points of Ref.~\cite{comment} for the discussion below. 

As mentioned in the former section, the study of the $np\to \pi^0\pi^0d$ and $np\to \pi^+\pi^-d$ reactions \cite{adlarson,adlardos} showed an unexpected narrow peak in the cross section around $2380$ MeV, which has been attributed to a dibaryon by the experimental team, branded $d^*(2380)$. It is interesting to remark that based on earlier data of the reaction, the peak had been attributed to a reaction mechanism based on sequential one pion production, $np\to \pi^-pp\to \pi^-\pi^+d$, together with $np\to\pi^+nn\to\pi^+\pi^-d$, in the work of Bar-Nir et al. \cite{barnir}. The second step, $pp\to\pi^+d$, was the object of theoretical investigation in \cite{brack,green,weise}, and was shown to be driven by $\Delta(1232)$ excitation. A reformulation of the idea of these works was done by us in Ref.~\cite{ikeno}, from a Feynman diagrammatic point of view, showing that the process developed a triangle singularity (TS) \cite{brack,landau,bayarguo}. This finding is relevant to the present discussion because it is well known that a TS gives rise to an Argand plot like the one of an ordinary resonance, even if the origin is a kinematical singularity and not the presence of a genuine physical state \cite{compass,guofeng}.

The idea of Bar-Nir et al. has been retaken in the former reaction and, making some reasonable approximations and using experimental data on $np(I=0)\to pp \pi^-$ and $pp\to \pi^+d$ reactions, a peak is obtained for the $pp\to\pi^+\pi^-d$ reaction in qualitative agreement with experiment in the position, width and strength. Even with the approximations done, and the qualitative agreement found, such agreement, together with the result of Bar-Nir et al., can hardly be an accident and offer an alternative explanation of the peak observed in the experiment.



Coming back to the manuscript of \cite{comment}, it  contains eight points that we reply here one by one.
\begin{itemize}
 \item[i)] In point 1) of the comment the authors complain about us enlarging the errors in the $pn (I=0) \to \pi^- pp$ reaction. The reason for that is that the cross section for this reaction is obtained using isospin symmetry with the relation 
\begin{equation}
\hspace{1cm}\sigma_{np(I=0) \to pp \pi^-} = ( \sigma_{np \to pp \pi^-} - \sigma_{pp \to pp \pi^0}/2)\ .\label{eq:sig}
\end{equation}

But the problem is that there are huge cancellations in this formula and the result is ten times smaller than each individual term. What we did is to assume a $5$\% uncertainty in the terms from isospin violation and determine the errors in the results. These are then systematic uncertainties that we think should have been considered by the experimentalists but they did not. So, we did. In the high energy of the spectrum, where the cross section falls down and produces the shape of the cross section, the systematic errors are much bigger than the statistical ones, hence it does not matter how the statistical ones are summed to them. The size is given by the systematic errors. In any case, the cross sections that we obtain have nothing to do with these errors.
 \item[ii)]  In point 2) the authors of the comment make a point about we not getting a precise description of the data. With the approximations that we did above we cannot pretend to get that precise agreement. We already consider an accomplishment that in such a complicated reaction we could get qualitatively a peak for the reaction at the right energy,  with the right width and the right strength at a qualitative level. 
 \item[iii)] The argument of this point is weak. First let us note that in our work of Ref.~\cite{ikeno} for the $pp\to \pi^+d$ reaction we proved the dominance of the $^1D_2$ as found experimentally in \cite{arndt,Oh:1997eq}. Second, the argument states that because in the $np(I=0)\to\pi^-pp$ reaction the invariant mass of $\pi^-p$ is big, then the one of $pp$ is small and only accommodates $L=0,L=1$ waves, not $D$-waves necessary for the overlap of the two-step mechanism. It is interesting to make this argument more quantitative. We have two situations where $M(pp)$ is easily evaluated. They correspond to the case of $M(\pi^-p)\vert_{\mathrm{min}}=m_p+m_{\pi^-}$ and $M(\pi^-p)\vert_{\mathrm{max}}=\sqrt{s}-m_p$. In the first case the $p$ and $\pi^-$ move together in opposite direction to the other proton. In the second case one proton is produced at rest. The $M(pp)$ is trivially evaluated in these two cases and for the energy $T_p=1200$~MeV of Fig.~6 of Ref.~\cite{wasasingle} we find,
 \begin{itemize}
 \item[a)] $M(pp)$ (at $M(\pi^-p)\vert_{\mathrm{min}}$) $=2239.47$~MeV, with an excess energy of the two protons of $362.9$~MeV.
 \item[b)] $M(pp)$ (at $M(\pi^-p)\vert_{\mathrm{max}})$ $=1920.2$~MeV, with an excess energy of $43.65$~MeV.
 \item[c)] There is another situation which also allows for an easy evaluation. This is at the peak of the distribution around $M(\pi^-p)=1370$~MeV where the $M(\pi^-p)$ from either of the protons is about the same and enhances the contribution in this region. There we have,
 $$\hspace{1cm} 2\,M^2(\pi^-p)+M^2(pp)=s+2\,m^2_p+m^2_{\pi^-}\ ,$$
 from where we get $M(pp)\simeq 1951$ MeV corresponding to about $75$~MeV excess energy for the two protons. Assuming relative distances of the produced two protons of $r\simeq 2.13$~fm, the radius of the deuteron, corresponding to the range of pion exchange, we find that the angular momentum, $L\sim r\times p$, can reach up to $L=6$ in the case of a), $L=2$ in the case b), and $L=3$ in the most favorable case corresponding to case c).
 \end{itemize}

  In the comment of \cite{comment}, $L=2$ is already ruled out and, without any calculation the sequential pion production cross section is deemed very small, contradicting the conclusions of \cite{barnir}. Actually, with the dominance of the Roper excitation, as claimed in \cite{comment}, we could already observe that $S=0$ for the two protons is the dominant mode in $np(I=0)\to\pi^-pp$, as in the second step of \cite{ikeno} for the $pp\to \pi^+d$ reaction, and several $L$ values are allowed. Work continues along these lines since Roper excitation is not the only ingredient of the $np(I=0)\to\pi^-pp$ reaction. 
 
  \item[iv)] Point 4) is illustrative.  In two independent papers, \cite{colin} and \cite{miguel}, it was shown that there was a relationship between the $pn (I=0) \to \pi^+ \pi^- d$ reaction and the one where the $np$ of the deuteron would become free states.  The existence of the peak of the  $pn (I=0) \to \pi^+ \pi^- d$ reaction had as a consequence a peak also in $pn  \to \pi^+ \pi^- n p$ and related reactions at the same energy. However, this was the case independently of which is the reason for the peak in the fusion reaction. This is a key point. Actually the authors of the comment,  in order to calculate the cross section of the open reactions, used the results of these references and added the contribution to the results of the standard model for these reactions that they also use from \cite{luisal}. Yet, they see that as an evidence of a dibaryon, while it was clearly shown in \cite{colin,miguel} that the new contribution was necessary whichever be the reason for the fusion reaction.
  \item[v)] In point 5) the comment complains that we do not calculate differential distributions. This is true, but we did not need that to prove our points. At the qualitative level that we worked, we showed that the distribution had to peak at small invariant masses of the two pions because we had two contributions: $pn (I=0) \to \pi^- pp$ followed by $pp \to \pi^+ d$, together with $pn (I=0) \to \pi^+ nn$ followed by $nn \to \pi^- d$. We could prove that when the momenta of the two pions are equal, the two amplitudes are identical and sum, producing a Bose enhancement.  In this case, the invariant mass of the two pions has its smallest value. This is why the cross section peaks at low $\pi \pi$ invariant mass. 
  \item[vi)] The comment claims that the picture of Bar-Nir \cite{barnir} presented in the former section, cannot explain the observed pole in $^3D_3-^3G_3$ $np$ partial waves. This statement is incorrect. The  $NN$, $I=0$, phase shifts will be affected by the peak in the  $pn (I=0) \to \pi^+ \pi^- d$ reaction because one can have $pn(I=0) \to \pi^+ \pi^- d \to  pn(I=0)$, where $\pi^+ \pi^- d$ is in an intermediate state and contributes to the inelasticities. This will be particularly the case in the quantum numbers preferred by the $pn (I=0) \to \pi^+ \pi^- d$ reaction which we discuss in our papers, in particular the $^3D_3$ partial wave, see discussion at the end of Ref.~\cite{ikeno}. At the energy of the peak of the $pn(I=0) \to \pi^+ \pi^- d$ reaction the  $np \to np$ amplitude will have an enhanced imaginary part due to the optical theorem and this has effects  in the phase shift at this energy. This can be said of most of the reactions claimed to see the dibaryon. What they see is a consequence of the peak seen in the $pn(I=0) \to \pi^+ \pi^- d$ reaction, whatever the reason for this peak be . This is the important point. The peak of the $pn(I=0) \to \pi^+ \pi^- d$ reaction will have repercussion in many observables, but this does not tell us that the reason for the peak has to be a dibaryon. Whatever the reason be, it will have consequences. Actually, the repercussion of the peak of the $pn(I=0)\to \pi^+\pi^-d$ reaction on the $^3D_3$ and $^3G_3$ partial waves was already discussed in \cite{colin,miguel}. Although it might look like the $pn\to \pi^+\pi^-d\to pn$ process would lead to a small contribution to the $ pn \to pn$ amplitude, the relatively large strength of the $pn \to \pi^+\pi^-d$ peak, makes this two step process not to small as was shown in \cite{colin,miguel}. It is also worth mentioning that small terms in an amplitude can show up more clearly in polarization observables than in direct cross sections, as shown in Ref.~\cite{luisroca}.
  
  The pole or resonant structure is guaranteed by the triangle singularity of the last step $pp\to\pi^+d$ shown in \cite{ikeno}. It is well-known that a triangle singularity produces an Argand plot similar to the one of a resonance \cite{compass, guofeng}.
\item[vii)] The comment of this point is again weak. It mentions that the ``$d^*(2380)$'' has been seen in the $\gamma d\to d\pi^0\pi^0$ and $\gamma d\to pn$ reactions. Actually in Refs. \cite{Ishikawa1,Ishikawa2} what one observes is a deviation of the experimental cross section from the theoretical calculations of \cite{fix,egorov} at low photon energies. These calculations are based on the impulse approximation and the $\pi$ rescattering terms are neglected. Actually, the rescattering contributions of pions are important, particularly at low energies because the momentum transfer is shared between two nucleons and one picks up smaller deuteron momentum components where the wave function is bigger. Thus, concluding that the discrepancies seen with experiment of a calculation based upon the impulse approximation are due to the dibaryon is an incorrect conclusion.
   
   Actually, we can add more to this discussion. The authors of Refs. \cite{Ishikawa1,Ishikawa2} also studied the $\gamma d\to \pi^0 \eta d$ reaction \cite{ishi1,ishi2}. This reaction was throughly studied theoretically in \cite{alber}, where it was found that the pion rescattering mechanisms were very important, and the most striking feature of the reaction, the shift of the shape of the invariant mass distributions was well reproduced. In the $\gamma d\to \pi^0\eta d$ reaction the $\eta$ rescattering had a small effect, only the $\pi$ rescattering was relevant. In the $\gamma d\to \pi^0\pi^0 d$ reaction the two pions can rescatter making the rescattering mechanism in $\gamma d\to \pi^0\pi^0 d$ even more important than in $\gamma d\to \pi^0\eta d$.
   
   The $\gamma d\to \pi^0\pi^0 d$ reaction has been measured more accurately in \cite{jude}. The same comments can be done concerning this work, since comparison with the data is done with the impulse approximation of Refs. \cite{fix,egorov}. Actually in that paper three dibaryons are claimed. It is not the purpose of this discussion to polemize with these conclusions but we cannot refrain from noting that a fit to the data with a straight line gives a better $\chi^2$ than the one with the three dibaryons.
   
   Concerning the $\gamma d\to p n$ (or $pn\to \gamma d$) signals observed in polarization observables in \cite{ikeda,bashkanov1,bashkanov2}, the following considerations are in order. The cross section for $\gamma d\to pn (pn\to \gamma d)$ has a clear peak due to the $\Delta(1232)$ excitation around $E_\gamma=260$ MeV \cite{rossi,whis}. This reaction is similar to the $pp\to \pi^+d$ reaction studied by us in Ref.~\cite{ikeno}, which develops a triangle singularity. It is easy to see using the same procedure as in \cite{ikeno} that the $\gamma d\to pn$ reaction is also driven by the same triangle singularity. In the cross section one does not see any trace of the ``$d^*(2380)$''. However, it is well known that polarization observables are sensitive to small terms of the amplitudes which do not show in integrated cross sections \cite{luisroca}. Thus, the combined reaction $np\to \pi^+\pi^-d\to \gamma d$ provides a contribution to the $np\to \gamma d$ reaction through an intermediate state which has a peak in the ``$d^*(2380)$'' region. As it is the case in \cite{luisroca}, this small amplitude can show up in polarization observables, justifying the observation of \cite{ikeda,bashkanov1,bashkanov2}. Yet, this cannot be seen as a proof of the existence of a dibaryon since it will occur whichever be the reason for the $pn(I=0)\to\pi^+\pi^-d$ peak.  
  \item[viii)] The point 8) is also illustrative. The authors of the comment claim that the cross section for the $pn (I=I) \to \pi^+ \pi^- d$  in our approach should be about $4$ times bigger than the one for $pn(I=0) \to \pi^+ \pi^- d$ , giving hand-waiving arguments, while experimentally it is about $10$ times smaller. Once again, this has a very easy explanation. As we have commented before, the two step process has two amplitudes: $pn (I=0) \to \pi^- pp$ followed by $pp \to \pi^+ d$, and $pn (I=0) \to  \pi^+ nn$ followed by $nn \to \pi^- d$. In the case of equal momenta of the pions the two amplitudes sum and produce an enhancement of the cross section.  On the contrary, for $I=1$ we have: $pn (I=1) \to \pi^- pp$ followed by $pp \to \pi^+ d$, and $pn (I=1) \to \pi^+ nn$ followed by $nn \to  \pi^- d$. But in this case, the two amplitudes cancel exactly. We proved that analytically, but it has to be like that because the two pions are in $I=1$ , which means p-wave and they cannot go together. 
   \item[ix)] Finally, the comment about the Argand plot has an easy answer, and has been mentioned before. Since the last step of our mechanism contains a triangle singularity, this creates a structure very similar to the one of a normal resonance, as discussed in the paper of the COMPASS collaboration \cite{compass}. 
  \end{itemize}
\section{Acknowledgments}
We thank Luis Alvarez Ruso for a careful reading of the paper and useful suggestions.
R. M. acknowledges support from the CIDEGENT program with Ref. CIDEGENT/2019/015 and from the spanish national grant PID2019-106080GB-C21. The work of N. I. was partly supported by JSPS Overseas Research Fellowships and JSPS KAKENHI Grant Number JP19K14709. This work is also partly supported by the Spanish Ministerio de Economia y Competitividad and European FEDER funds under Contracts No. FIS2017-84038-C2-1-P B and No.  FIS2017-84038-C2-2-P B. This project has received funding from the European Union’s Horizon 2020 research and innovation programme under grant agreement No. 824093 for the STRONG-2020 project.

\bibliography{biblio}

\end{document}